\documentclass[a4paper]{jpconf}

\usepackage{citesort}
\usepackage{graphicx}

\begin{document}

\title{Status of FNAL SciBooNE experiment}

\author{Yasuhiro Nakajima, for the SciBooNE collaboration}

\address{Department of Physics, Kyoto University, Kyoto 606-8502, Japan}

\ead{nakajima@scphys.kyoto-u.ac.jp}

\begin{abstract}
SciBooNE is a new experiment at FNAL which will make precision neutrino-nucleus cross
section measurements in the one GeV region. These measurements are essential for the future neutrino oscillation experiments.
We started data taking in the antineutrino mode
on June 8, 2007, and collected $5.19\times 10^{19}$ protons on target (POT)  before the accelerator shutdown in August.
The first data from SciBooNE are reported in this article.

\end{abstract}

\section{Introduction}

Neutrino oscillation is a tool to study the nature of neutrino mass and mixing, and is a probe of physics at high energy scales beyond the standard model.
In future accelerator neutrino experiments, precise knowledge of neutrino cross sections with nuclei are important to achieve the best sensitivity.
On the other hand, the typical accuracy of neutrino-nucleus cross sections in the one GeV region, which is relevant to many future experiments, is $\sim$20\% with our current knowledge.

The SciBooNE experiment (FNAL E954) \cite{AguilarArevalo:2006se} is designed to measure
the neutrino cross sections on carbon in the one GeV region.
SciBooNE's neutrino beam energy is well matched to T2K\cite{t2k}. Additionally, SciBooNE serves as a near detector for MiniBooNE\cite{AguilarArevalo:2007it} by constraining the neutrino fluxes.

\section{Experimental setup}
\subsection{Booster neutrino beam at FNAL}

The primary proton beam, with kinetic energy 8~GeV, is extracted from the Fermilab
Booster and steered to strike a 71 cm long, 1 cm diameter beryllium target. The proton
beam typically has $4\times 10^{12}$ protons per $\sim$1.6~$\mu$sec beam spill.
The mesons, primarily $\pi^+$, generated by the $p\rm{-Be}$ interactions are focused with a magnetic horn and decay in the following 50m decay volume, producing an intense neutrino beam with the peak energy of 0.7 GeV.
When the horn polarity is reversed,
$\pi^-$ are focused and hence a predominantly antineutrino beam is created.

We plan to run $10^{20}$ POT each in neutrino mode and in antineutrino mode.
Since the horn was operating in antineutrino mode for MiniBooNE, we started
running with antineutrino beam.

\subsection{SciBooNE detector}

The SciBooNE detector is located 100 m downstream from the beryllium target, on the axis of the neutrino beam direction.
The detector complex consists of three sub-detectors: a fully active fine grained
scintillator tracking detector (SciBar), an electromagnetic calorimeter (EC) and
a muon range detector (MRD).

The SciBar detector was originally built and operated in the K2K experiment\cite{Nitta:2004nt}. After the K2K was completed, the SciBar was shipped and reassembled at FNAL together with the EC which was used in the CHORUS\cite{Buontempo:1995qe} and K2K experiments.
The MRD is a newly constructed hodoscope made with recycled iron plates and scintillators from past FNAL experiments. 

The SciBar consists of 14,336 extruded plastic scintillator strips,
each with dimension of $1.3\times 2.5\times 300$ cm$^3$. The scintillators are arranged vertically
and horizontally to construct a  $3\times 3\times 1.7$ m$^3$ detector. 
The detector itself is the neutrino target and can reconstruct all the charged particles from a neutrino interaction together with particle identification capability using dE/dx.

The EC is installed downstream of the SciBar,
and is made of scintillating fibers embedded in lead foil.
 It has a thickness of 11 radiation length along the beam direction to measure
$\pi^0$ produced in neutrino interactions and the intrinsic $\nu_e$ contamination.
The energy resolution is $14\%/\sqrt{E\rm{[GeV]}}$.
The MRD is located downstream of the EC in order to measure the momentum of muons up to
1.2 GeV$/c$ using the muon range. It consists of 12 layers of 2''-thick iron plates sandwiched
between layers of 6 mm-thick plastic scintillator planes.

\section{Detector commissioning}

The detector construction was completed in March 2007.
Immediately after construction, the detectors were commissioned with cosmic ray muons.
In April 2007, the detectors were moved and installed in the experimental hall on the
neutrino beam line. The second commissioning using the antineutrino beam started
in May 2007, and we successfully detected events on May 29th, 2007.

\begin{figure}[htbp]
  \includegraphics[keepaspectratio=true,width=18pc]{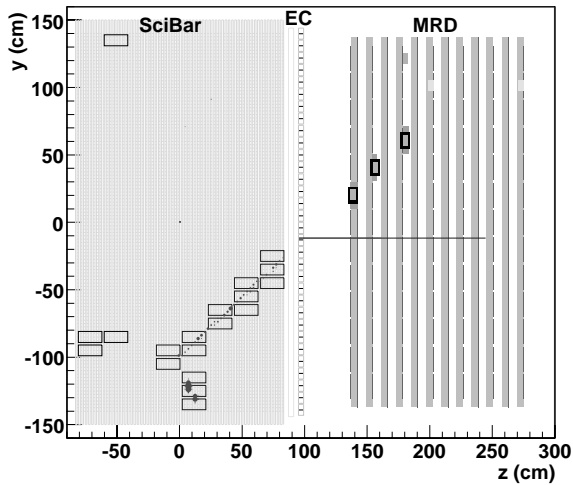}
  \hspace{1.5pc}
  \begin{minipage}[b]{18pc}\caption{A candidate of an antineutrino charged current 
  	quasi-elastic event
            ($\bar{\nu}_\mu + p \to \mu^+ + n$).
            Circles and boxes on the SciBar indicate hits of ADC and TDC respectively.
            The area of the circle is proportional to the ADC value.
            For the EC, the size of a bar corresponds to the energy deposit.
            Finally, framed boxes on the MRD indicate TDC hits within the beam-timing.}
  \label{fig:event_display_antinu}
  \end{minipage}
\end{figure}

Figure~\ref{fig:event_display_antinu} shows a display of
an antineutrino charged current quasi-elastic event ($\bar{\nu}_\mu + p \to \mu^+ + n$) candidate from the commissioning period.
The longer track penetrating through the SciBar and the EC to the MRD is a muon, and the short
track with large energy deposit in the SciBar is presumed to be a proton recoiled by
a neutron from the vertex of neutrino interaction.

\section{First antineutrino data taking}

After commissioning, we took data in antineutrino mode from June 8, 2007 
to the accelerator shutdown in August 2007.
We collected $5.19\times 10^{19}$ POT in two months, which is about half of
our goal for antineutrino data.

In order to confirm the stability of the detector and the beam, we select charged current (CC) candidate events by requiring a track exiting from the SciBar whose vertex is inside the fiducial volume of the SciBar, corresponding to a mass of 10.6 tons.

\begin{figure}[htbp]
\begin{minipage}[t]{18pc}
  \includegraphics[keepaspectratio=true,width=18pc]{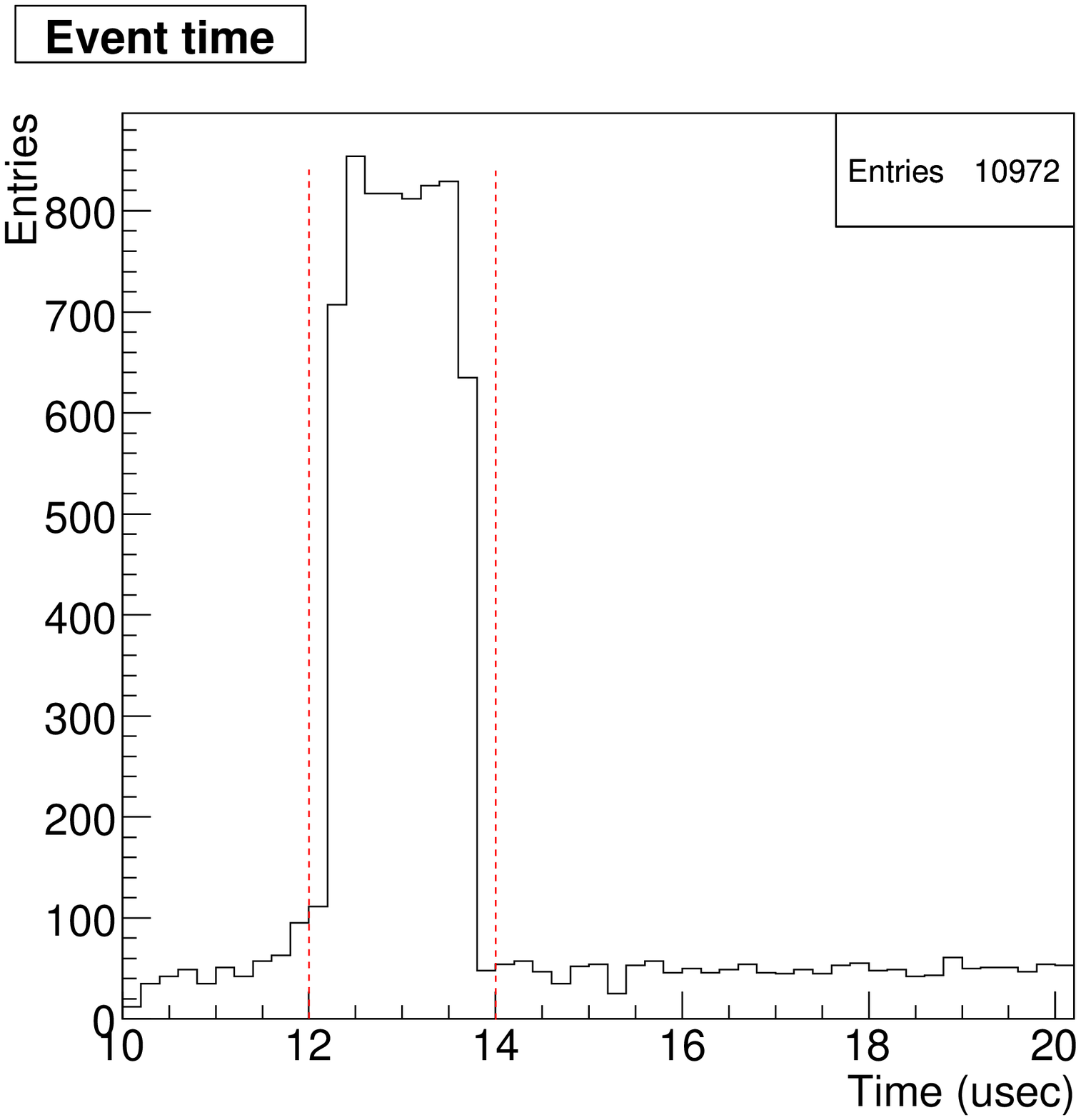}
  \caption{Timing distribution of charged current candidate events with the vertices inside the SciBar.
            Dashed lines indicate the 2 $\mu$sec beam-timing window.}
  \label{fig:scibar_eventtime}
\end{minipage}
\hspace{1.5pc}
\begin{minipage}[t]{18pc}
  \includegraphics[keepaspectratio=true,width=18pc]{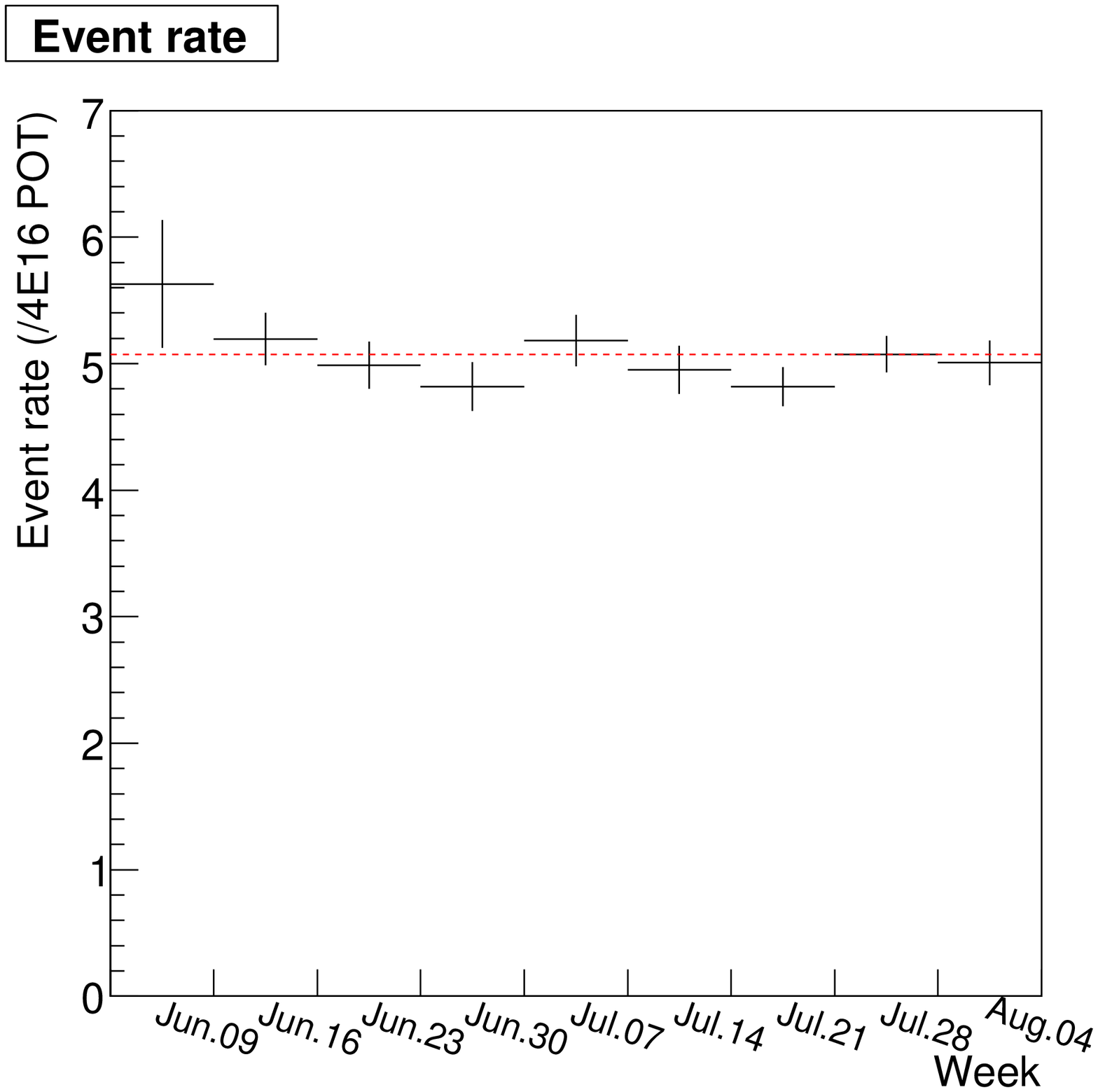}
  \caption{Event rate of the CC candidate events in the beam timing.
            The rate is normalized by the number of protons on target. The error bars show the 	statistical errors.
            Dashed lines indicate the average rate.}
  \label{fig:scibar_eventrate}
\end{minipage}
\end{figure}

Figure~\ref{fig:scibar_eventtime} shows the timing distribution of the CC
candidate events. One can clearly
see events clustered in the beam-timing window. Figure~\ref{fig:scibar_eventrate}
shows the number of events after cosmic ray background subtraction per week per $4\times 10^{16}$ POT in the beam-timing window
without any acceptance corrections. 
The number of cosmic ray background events is estimated from the events outside the beam-timing window.
The rate is stable within the statistical error
for the whole period of the running.

\section{Summary and Future prospects}
The SciBooNE data taking successfully started in antineutrino mode running in June 2007.
We confirm that the detector and beam ran stably for more than 2 months.

After the 2007 summer shutdown, we started taking data in neutrino mode and 
plan to collect
$10^{20}$ POT by spring of 2008. After the neutrino mode run, we will switch
back to antineutrino mode and collect the additional $5\times 10^{19}$ POT data
before summer 2008.

\ack
 SciBooNE collaboration gratefully acknowledges the support from various
 grants and contracts from the Department of Energy (U.S.), the National
 Science Foundation (U.S.), the MEXT (Japan), the INFN (Italy) and the
 Spanish Ministry of Education and Science.
 The author was supported by Japan Society for the Promotion of Science.

\section*{References}
\bibliography{ref}

\providecommand{\newblock}{}
\begin{thebibliography}{1}
\expandafter\ifx\csname url\endcsname\relax
  \def\url#1{{\tt #1}}\fi
\expandafter\ifx\csname urlprefix\endcsname\relax\def\urlprefix{URL }\fi
\providecommand{\eprint}[2][]{\url{#2}}

\bibitem{AguilarArevalo:2006se}
Aguilar-Arevalo A~A {\em et~al.\/} 2006  (\textit{Preprint}
  \eprint{hep-ex/060122})

\bibitem{t2k}
Itow Y {\em et~al.\/} 2001  (\textit{Preprint} \eprint{hep-ex/0106019})

\bibitem{AguilarArevalo:2007it}
Aguilar-Arevalo A~A {\em et~al.\/} 2007 {\em Phys. Rev. Lett.\/} {\bf 98}
  231801 (\textit{Preprint} \eprint{arXive:0704.1500 [hep-ex]})

\bibitem{Nitta:2004nt}
Nitta K {\em et~al.\/} 2004 {\em Nucl. Instrument. Meth.\/} A {\bf 535} 147

\bibitem{Buontempo:1995qe}
Buontempo S {\em et~al.\/} 1995 {\em Nucl. Phys. Proc. Suppl.\/} {\bf 44} 45

\end{thebibliography}

\end{document}